\documentclass[12pt]{article}
\usepackage{pstricks}
\usepackage{color}
\usepackage{cite}
\usepackage{array}
\usepackage{epsfig}
\usepackage{amssymb}
\usepackage{graphics,graphpap}
\usepackage{amssymb}
\usepackage{amsmath}
\usepackage{slashed}
\usepackage{dsfont}
\usepackage{a4wide}

\def\eps{\varepsilon}
\def\epe{\varepsilon'/\varepsilon}

\newcommand{\gev}{\, {\rm GeV}}
\newcommand{\mev}{\, {\rm MeV}}

\newcommand{\mt}{m_{\rm t}}
\newcommand{\mtb}{\overline{m}_{\rm t}}

\newcommand{\be}{\begin{equation}}
\newcommand{\ee}{\end{equation}}
\newcommand{\bea}{\begin{eqnarray}}
\newcommand{\eea}{\end{eqnarray}}

\newcommand{\bi}{\begin{itemize}}
\newcommand{\ei}{\end{itemize}}
\newcommand{\ord}{{\cal O}}

\newcommand{\vcb}{|V_{cb}|}
\newcommand{\vtd}{|V_{td}|}
\newcommand{\vub}{|V_{ub}|}
\newcommand{\vts}{|V_{ts}|}
\newcommand{\vus}{|V_{us}|}

\def\kpn{K^+\rightarrow\pi^+\nu\bar\nu}

\def\klpn{K_{L}\rightarrow\pi^0\nu\bar\nu}

\usepackage{graphicx}

\medskipamount 

 \def\s#1{\setbox0=\hbox{$#1$}%
   \rlap{\ifdim\wd0>.7em\kern.22\wd0\else\kern.1\wd0\fi /}#1}


\begin{document}

\begin{titlepage}
\begin{flushright}
\begin{tabular}{l}
TTP16-006 \\
{FLAVOUR(267104)-ERC-117}
\end{tabular}
\end{flushright}
\vskip1.2cm
\begin{center}
{\Large \bf \boldmath
Universal Unitarity Triangle 2016 and the Tension \vspace{2mm}\\
Between $\Delta M_{s,d}$ and $\varepsilon_K$  in CMFV Models}
\vskip1.0cm
{\bf \large
Monika Blanke$^{a,b}$ and Andrzej J. Buras$^c$}
\vskip0.3cm
$^a$ {Institut fur Kernphysik, Karlsruhe Institute of Technology,
  Hermann-von-Helmholtz-Platz 1,
  D-76344 Eggenstein-Leopoldshafen, Germany}\vspace{1mm}\\
$^b$ {Institut fur Theoretische Teilchenphysik,
  Karlsruhe Institute of Technology, Engesserstra\ss e 7,
  D-76128 Karlsruhe, Germany}\vspace{1mm}\\
$^c$ TUM-IAS, Lichtenbergstr. 2a, D-85748 Garching, Germany\\
Physik Department, TUM, D-85748 Garching, Germany\\

\vskip0.51cm


\vskip0.35cm

{\large\bf Abstract\\[10pt]} \parbox[t]{\textwidth}{
Motivated by the recently improved results from the Fermilab Lattice and MILC Collaborations  on 
the hadronic matrix elements entering  $\Delta M_{s,d}$ in
 $B_{s,d}^0-\bar B_{s,d}^0$ mixing, we determine the Universal Unitarity Triangle (UUT) in models with Constrained Minimal 
Flavour Violation (CMFV). Of particular importance are the very  precise determinations  of the ratio {$\vub/\vcb=0.0864\pm0.0025$} and of the angle {$\gamma=(63.0\pm 2.1)^\circ$.} They follow in this framework from
the experimental values of $\Delta M_{d}/\Delta M_s$ and of the CP-asymmetry $S_{\psi K_S}$. 
As in CMFV models the new contributions to meson mixings can be described by a single flavour-universal variable $S(v)$, we next determine the CKM matrix elements $\vts$,  $\vtd$,  $\vcb$ and $\vub$ as functions of $S(v)$ using the experimental value of $\Delta M_s$ as input. The lower bound on $S(v)$ in these models, derived by us in 2006, implies then  {\it upper} bounds on these four CKM elements and on the CP-violating parameter $\varepsilon_K$, which turns out to be
 significantly below its experimental value. This strategy avoids the 
use of tree-level determinations of $\vub$ and $\vcb$ that are presently 
subject to {considerable} uncertainties. 
 On the other hand if $\varepsilon_K$ is used instead of 
$\Delta M_s$  as input,  $\Delta M_{s,d}$  are found significantly above the data.
In this manner we point out that the new lattice data have significantly sharpened  
the tension between $\Delta M_{s,d}$ and $\varepsilon_K$ within the CMFV framework. {This implies} the presence of new physics contributions beyond this framework that are responsible for the breakdown of the flavour universality of the function $S(v)$. We also present the implications of these results for $\kpn$, $\klpn$ and $B_{s,d}\to\mu^+\mu^-$ within the {Standard Model}. 
}

\end{center}
\end{titlepage}

\setcounter{footnote}{0}

\newpage

\section{Introduction}
\label{sec:1}
Already for decades the $\Delta F=2$ transitions in the 
down-quark sector, that is $B^0_{s,d}-\bar B^0_{s,d}$ and 
$K^0-\bar K^0$ mixings, have been vital in constraining the {Standard Model (SM)} and in the search 
for new physics (NP) \cite{Isidori:2010kg,Buras:2013ooa}. However, theoretical uncertainties related to 
the hadronic matrix elements entering these transitions and their large sensitivity to the
CKM parameters {so far} precluded clear cut conclusions about the presence of 
new physics (NP). 

The five observables of interest are 
\be\label{great5}
\Delta M_s,\quad \Delta M_d, \quad S_{\psi K_S},\quad S_{\psi \phi}, \quad \varepsilon_K
\ee
with $\Delta M_{s,d}$ being the mass differences in 
$B^0_{s,d}-\bar B^0_{s,d}$ mixings and $S_{\psi K_S}$ and $S_{\psi \phi}$ 
the corresponding mixing induced CP-asymmetries.
$\varepsilon_K$ describes the size of the indirect CP violation in $K^0-\bar K^0$ mixing. $\Delta M_{s,d}$ and  $\varepsilon_K$ 
are already known with impressive precision. The asymmetries  $S_{\psi K_S}$ and $S_{\psi \phi}$ are less precisely  measured but have the advantage of being subject to only very small hadronic uncertainties. {We do not include $\Delta M_K$ in (\ref{great5}) as it is subject to much larger theoretical uncertainties than the 
five observables in question.}

The hadronic uncertainties in   $\Delta M_{s,d}$ and $\varepsilon_K$  within 
the SM and CMFV models reside within a good approximation in the parameters
\be\label{hpar}
 F_{B_s}\sqrt{\hat B_{B_s}},\quad  F_{B_d} \sqrt{\hat B_{B_d}}, \quad \hat B_K\,.
\ee
Fortunately, during the last years these uncertainties decreased significantly.
In particular, concerning  $F_{B_s}\sqrt{\hat B_{B_s}}$ and $F_{B_d} \sqrt{\hat B_{B_d}}$,
an impressive progress has recently been made by the Fermilab Lattice and MILC Collaborations (Fermilab-MILC) 
 that find \cite{Bazavov:2016nty}
\be\label{Kronfeld}
 F_{B_s}\sqrt{\hat B_{B_s}}=(274.6\pm8.8)\mev,\qquad  F_{B_d} \sqrt{\hat B_{B_d}}=
(227.7\pm 9.8)\mev \,,
\ee
with uncertainties of $3\%$ and $4\%$, respectively. An even higher 
precision is achieved for the ratio 
\be\label{xi}
\xi=\frac{F_{B_s}\sqrt{\hat B_{B_s}}}{F_{B_d}\sqrt{\hat B_{B_d}}}=1.206\pm0.019\,.
\ee
This value is significantly lower than the central value $1.27$ in the previous lattice estimates \cite{Aoki:2013ldr} and its reduced uncertainty by a factor of three plays an  important  role in our analysis.
{The ETM Collaboration has also presented results for matrix elements of all five operators entering $B_{d,s}-\bar B_{d,s}$ mixing \cite{Carrasco:2013zta}.
This work however only employs two flavours of sea quarks and does not estimate the uncertainty from quenching the strange quark.
The ETM and Fermilab-MILC results for matrix elements differ by $\sim5\%$, or $\sim1\sigma$, which could arise from the omitted strange sea.
We think it is safer to avoid this issue and use only the Fermilab-MILC results with $N_f=2+1$. However we note that the result for $\xi$ obtained  by the ETM collaboration supports a rather low value of $\gamma$ from the universal unitarity triangle (UUT).}
 {An extensive list of} references to other
 lattice determinations of these parameters can be found in \cite{Bazavov:2016nty}.

Lattice QCD also made an impressive progress in
the determination of the parameter $\hat B_K$ which enters the evaluation of
$\varepsilon_K$ \cite{Aoki:2010pe,Bae:2010ki,Constantinou:2010qv,Colangelo:2010et,Bailey:2012bh,Durr:2011ap}.  The most recent preliminary world 
average from FLAG reads $\hat B_K=0.7627(97)$ \cite{Vladikas:2015bra}, 
 very close to its large $N$ 
value $\hat B_K=0.75$ \cite{Gaiser:1980gx,Buras:1985yx}.
 Moreover the analyses in \cite{Gerard:2010jt,Buras:2014maa} show that 
$\hat B_K$ cannot be larger than $0.75$  but close to it.
Taking the 
present results and  precision of lattice QCD into account it is then a 
{good approximation to set $\hat B_K=0.750 \pm 0.015$.
In the evaluation of $\varepsilon_K$ we also take into account long distance contributions parametrised by $\kappa_\varepsilon = 0.94\pm 0.02$ \cite{Buras:2010pza}. Note that at present the theoretical uncertainty in $\eps_K$  is dominated by the parameter $\eta_{cc} = 1.87 \pm 0.76$ \cite{Brod:2011ty} summarising NLO and NNLO QCD corrections to the charm quark contribution. We take these uncertainties into account.}

With $\vus$ determined already very precisely, the main uncertainties in 
the CKM parameters reside in  
\be\label{par}
\vcb, \qquad \vub, \qquad \gamma,
\ee
with $\gamma$ being one of the angles of the unitarity triangle (UT). These 
three parameters can be determined from tree-level decays that are 
subject to only very small  NP contributions.
However the tensions between inclusive and exclusive determinations of $\vub$ and 
{to a lesser extent of} $\vcb$  do not yet allow for clear cut conclusions on their values. Moreover,
the current direct measurement of $\gamma$ is not 
precise~\cite{LHCb-CONF-2016-001} 
\begin{equation}\label{gamma}
    \gamma = (70.9^{+7.1}_{-8.5})^\circ.
\end{equation}
This is consistent with $\gamma$ from the U-spin analysis of $B_s\to K^+K^-$ and $B_d\to\pi^+\pi^-$ decays ($\gamma=(68.2\pm 7.1)^\circ$) \cite{Fleischer:2010ib}. {The U-spin analysis by LHCb \cite{Aaij:2014xba} on the other hand finds a lower value $ \gamma = ( 63.5^{+7.2}_{-6.7})^\circ$ in good agreement with the result from the UUT analysis in \eqref{newgamma}.}

The present uncertainties in $\vub/\vcb$ and $\gamma$ from tree-level decays preclude then a precise determination of the so-called {\it reference unitarity triangle} (RUT) \cite{Goto:1995hj} which is expected to be practically independent of the presence of NP. 
In addition the uncertainty in $\vcb$ prevents precise predictions for 
$\varepsilon_K$ and $\Delta M_{s,d}$ in the SM. However in the SM 
and more generally models with constrained minimal flavour violation (CMFV) \cite{Buras:2000dm,Buras:2003jf,Blanke:2006ig} it is possible to construct the so-called {\it universal unitarity triangle} (UUT) \cite{Buras:2000dm} for which the knowledge of 
$\vub/\vcb$ and $\gamma$ is not required. The UUT can be constructed from 
\be\label{UUT}
\frac{\Delta M_d}{\Delta M_s}, \qquad S_{\psi K_S}
\ee
and this in turn allows to determine $\vub/\vcb$ and $\gamma$.

The important virtue of this determination is its universality within 
CMFV models. In the case of $\Delta F=2$ transitions in the 
down-quark sector various CMFV models can only be distinguished by the value of a
single flavour universal real one-loop  function, the 
box diagram function $S(v)$, with $v$ collectively denoting the parameters of a given 
CMFV model. This function enters universally $\varepsilon_K$, $\Delta M_s$ 
and $\Delta M_d$ and cancels out in the ratio in (\ref{UUT}). Therefore 
the resulting UUT is the same in 
all CMFV models. Moreover it can be shown that in these models $S(v)$ is bounded from below by its SM value \cite{Blanke:2006yh}
\be\label{BBBOUND}
S(v)\ge S_0(x_t)= 2.32
\ee
with $S_0(x_t)$ given in (\ref{S0}). 

The recent results in (\ref{Kronfeld}) and (\ref{xi}) have a profound impact 
on the determination of the UUT.
 The UUT can be determined very precisely from the measured values of $\Delta M_d/\Delta M_s$ and $S_{\psi K_S}$. 
This in turn implies a precise knowledge of the ratio $\vub/\vcb$
and  the angle $\gamma$, both to be compared with their tree-level determinations. Also the side $R_t$ of the 
UUT can be {determined precisely} in view of the result for $\xi$ in (\ref{xi}).

In order to complete the  determination of the full CKM matrix without the use of any  tree-level determinations, except for $\vus$,  we will use two strategies:
\begin{itemize}
\item[{\bf\boldmath $S_1$:}]
{\boldmath $\Delta M_s$ \unboldmath} {\bf strategy}  in which the experimental value of 
$\Delta M_s$ is used to determine $\vcb$ as a function of $S(v)$, and $\varepsilon_K$ is then a derived quantity. 
\item[{\bf\boldmath $S_2$:}]
{\boldmath $\varepsilon_K$ \unboldmath}{\bf strategy} in which the experimental value of 
$\varepsilon_K$ is used, while $\Delta M_{s}$ is then a derived quantity and 
$\Delta M_d$ follows from the determined UUT. 
\end{itemize}

Both strategies use the determination of the UUT {by means of (\ref{UUT})}
 and allow 
to determine the whole CKM matrix, in particular $\vts$, $\vtd$, $\vub$ and $\vcb$ as functions of $S(v)$. Yet their outcome is very different, 
which signals the tension between $\Delta M_{s,d}$ and $\varepsilon_K$ 
in this framework. As we will demonstrate below, this tension, known already 
from previous studies \cite{Buras:2011wi,Buras:2012ts}, has been sharpened significantly through 
the results in (\ref{Kronfeld}) and (\ref{xi}). Using these two strategies 
separately allows to exhibit this tension transparently.
Indeed
\begin{itemize}
\item
The lower bound in (\ref{BBBOUND}) implies in $S_1$ {\it upper bounds} on 
 $\vts$, $\vtd$, $\vub$ and $\vcb$ which are saturated in the SM, and 
in turn allows to derive an {\it upper bound} on $\varepsilon_K$ in CMFV models 
that is saturated in the SM but turns out to be significantly {\it below} the data.
\item
The lower bound in (\ref{BBBOUND}) implies in $S_2$  also {\it upper bounds} on 
 $\vts$, $\vtd$, $\vub$ and $\vcb$ which are saturated in the SM. {However} 
the $S(v)$ dependence of these elements determined in this manner differs from
the one obtained in $S_1$, which in turn allows to derive  {\it lower bounds} on $\Delta M_{s,d}$ in CMFV models 
that are reached in the SM but turn out to be significantly {\it above} the data.
\end{itemize}

It has been known since 2008 that the SM experiences 
some tension in the correlation between $S_{\psi K_S}$ and $\varepsilon_K$  \cite{Lunghi:2008aa,Buras:2008nn,Bona:2009cj,Lenz:2010gu,Lunghi:2010gv}. 
 It should be emphasized that in CMFV models only 
the version of this tension in \cite{Buras:2008nn}, i.\,e.\ NP in $\varepsilon_K$, is possible as in these models there are no new CP-violating phases. Therefore $S_{\psi K_S}$ has to be used to determine the 
sole phase in these models, the angle $\beta$ in the UT, or equivalently the CKM phase, through the unitarity of the CKM matrix. The resulting low value of $\varepsilon_K$
can be naturally raised in CMFV models by enhancing the value of $S(v)$ 
or/and increasing the value of $\vcb$. 
However, as pointed out in \cite{Buras:2011wi,Buras:2012ts}, this spoils 
the agreement of the SM with the data on $\Delta M_{s,d}$, signalling
 the  tension between $\Delta M_{s,d}$ and $\varepsilon_K$ in CMFV models. The 
 2013  analysis of this tension in \cite{Buras:2013raa} found that 
 the situation of CMFV with respect to $\Delta F=2$ transitions would 
improve if more precise results for  $F_{B_s}\sqrt{\hat B_{B_s}}$ and $F_{B_d} \sqrt{\hat B_{B_d}}$ turned out to be lower than the values known in the 
spring of 2013. The recent results from \cite{Bazavov:2016nty} in (\ref{Kronfeld})
show the opposite. Both  $F_{B_s}\sqrt{\hat B_{B_s}}$ and   $F_{B_d} \sqrt{\hat B_{B_d}}$ increased. Moreover the more precise and significantly smaller value of  $\xi$ enlarges the tension in question.

 In view of the new lattice results, in this paper we take another look 
at CMFV models. Having more precise values for $F_{B_s}\sqrt{\hat B_{B_s}}$,  $F_{B_d} \sqrt{\hat B_{B_d}}$ and  $\xi$ than in 2013, our strategy outlined 
above differs from the one in \cite{Buras:2013raa}. In particular we take $\gamma$ to be a derived quantity and not an input as done in the latter paper. Moreover, we will be able to reach much  firmer conclusions than it was possible in 2013. In particular, in contrast to \cite{Buras:2013raa} and also to \cite{Bazavov:2016nty} at no place in 
our paper tree-level determinations of $\vub$, $\vcb$ and $\gamma$ are used. 
However we compare our results with them.

It should be mentioned that Fermilab-MILC identified a significant tension 
between their results for the $B^0_{s,d}-\bar B^0_{s,d}$ mass differences and the tree-level determination of the CKM matrix within the SM. Complementary to their findings, we identify a significant tension within $\Delta F = 2$ processes, that is between $\varepsilon_K$ and 
$\Delta M_{s,d}$ in the whole class of CMFV models. Moreover, we determine 
very precisely the UUT, in particular the angle $\gamma$ in this triangle and 
the ratio $\vub/\vcb$, both valid also in the SM.

Our paper is organized as follows. In Section~\ref{sec:2} we determine first 
the UUT as outlined above, that in 2016 is significantly better known than 
in 2006 \cite{Blanke:2006ig} and in particular in 2000, when the UUT was 
first suggested \cite{Buras:2000dm}. Subsequently we execute the strategies $S_1$ and 
$S_2$ defined above. The values of  $\vts$,  $\vtd$,  $\vcb$ and $\vub$, 
resulting from these two strategies, differ significantly from each other 
which is the consequence of the tension between $\varepsilon_K$ and $\Delta M_{s,d}$ in 
question. In Section~\ref{sec:3} we 
present the implications of these results for $\klpn$, $\kpn$ and $B_{s,d}\to\mu^+\mu^-$ within the SM, obtaining again rather different results in $S_1$ and 
$S_2$.
In Section~\ref{sec:4} we briefly discuss how the $U(2)^3$ models 
face the new lattice data and comment briefly on other models. 
 We  conclude in Section~\ref{sec:5}.

\section{Deriving the UUT and the CKM}\label{sec:2}

\subsection{Determination of the UUT}

We begin with the determination of the UUT.
 For the mass differences in the $B^0_{s,d}-\bar B^0_{s,d}$ systems we have 
{the very accurate expressions}
\bea\label{DMS}
\Delta M_{s}&=&
17.757/{\rm ps}\cdot\left[ 
\frac{\sqrt{\hat B_{B_s}}F_{B_s}}{274.6\mev}\right]^2
\left[\frac{S(v)}{2.322}\right]
\left[\frac{\vts}{0.0390} \right]^2
\left[\frac{\eta_B}{0.5521}\right] \,,\\
\label{DMD}
\Delta M_d&=&
0.5055/{\rm ps}\cdot\left[ 
\frac{\sqrt{\hat B_{B_d}}F_{B_d}}{227.7\mev}\right]^2
\left[\frac{S(v)}{2.322}\right]
\left[\frac{\vtd}{8.00\cdot10^{-3}} \right]^2 
\left[\frac{\eta_B}{0.5521}\right]\,. 
\eea
The value $2.322$ in the normalization of $S(v)$ is its SM value for 
$m_t(m_t)=163.5\gev$ {obtained from}
\be\label{S0}
S_0(x_t)  = \frac{4x_t - 11 x_t^2 + x_t^3}{4(1-x_t)^2}-\frac{3 x_t^2\log x_t}{2
(1-x_t)^3}= 2.322 \left[\frac{\mtb(\mt)}{163.5\gev}\right]^{1.52}\,,
\ee
and $\eta_B$ is the perturbative QCD correction \cite{Buras:1990fn}. Our input parameters, equal 
to the ones used in \cite{Bazavov:2016nty}, are collected in Table~\ref{tab:input}.

\begin{table}[!tb]
\center{\begin{tabular}{|l|l|}
\hline
$m_{B_s} = 5366.8(2)\mev$\hfill\cite{Agashe:2014kda}	&  $m_{B_d}=5279.58(17)\mev$\hfill\cite{Agashe:2014kda}\\
 $\Delta M_s = 17.757(21) \,\text{ps}^{-1}$\hfill\cite{Amhis:2014hma}	&  $\Delta M_d = 0.5055(20) \,\text{ps}^{-1}$\hfill\cite{Amhis:2014hma}\\ 
$S_{\psi K_S}= 0.691(17)$\hfill\cite{Amhis:2014hma}
		&  $S_{\psi\phi}= 0.015(35)$\hfill \cite{Amhis:2014hma}\\
	$|V_{us}|=0.2253(8)$\hfill\cite{Agashe:2014kda} &
 $|\eps_K|= 2.228(11)\cdot 10^{-3}$\hfill\cite{Agashe:2014kda}\\
$F_{B_s}$ = $228.6(3.8)\mev$ \hfill \cite{Rosner:2015wva} & $F_{B_d}$ = $193.6(4.2)\mev$ \hfill \cite{Rosner:2015wva}  \\
$m_t(m_t)=163.53(85)\gev$ & $S_0(x_t)=2.322(18)$ \\
$\eta_{cc}=1.87(76)$\hfill\cite{Brod:2011ty} & $\eta_{ct}= 0.496(47)$\hfill\cite{Brod:2010mj}\\
$\eta_{tt}=0.5765(65)$\hfill\cite{Buras:1990fn}	&
$\eta_B=0.55(1)$\hfill\cite{Buras:1990fn,Urban:1997gw}\\
$\tau_{B_s}= 1.510(5)\,\text{ps}$\hfill\cite{Amhis:2014hma} &  $\Delta\Gamma_s/\Gamma_s=0.124(9)$\hfill\cite{Amhis:2014hma} \\
$\tau_{B_d}= 1.520(4)\,\text{ps}$\hfill\cite{Amhis:2014hma} &  $\kappa_\varepsilon = 0.94(2)$\hfill \cite{Buras:2010pza}
\\	       
\hline
\end{tabular}  }
\caption {\textit{Values of the experimental and theoretical
    quantities used as input parameters. For future 
updates see PDG \cite{Agashe:2014kda}  and HFAG \cite{Amhis:2014hma}. 
}}
\label{tab:input}
\end{table}

From (\ref{DMS}) and (\ref{DMD}) we find using (\ref{xi}) 
\be\label{DMSDMD}
\frac{\vtd}{\vts}=\xi\sqrt{\frac{m_{B_s}}{m_{B_d}}}\sqrt{\frac{\Delta M_d}{\Delta M_s}}= {0.2052\pm 0.0033}\,,
\ee
which perfectly agrees with \cite{Bazavov:2016nty}. The tree-level determination of this ratio, quoted in the latter paper and obtained from  {CKMfitter \cite{Charles:2015gya},
reads} 
\be\label{treeDMSDMD}
\frac{\vtd_\text{tree}}{\vts_\text{tree}}={0.2180\pm 0.0051}\,.
\ee
It is significantly higher than the value in (\ref{DMSDMD}). It should be 
emphasized that {the values of} $\vcb$ and $\vub$ to a very good approximation do not enter this ratio. Therefore this discrepancy is not a consequence of the tree-level determinations of $\vcb$ and $\vub$. As we will demonstrate below it is the consequence of the value of the angle $\gamma$, which due to the small value of $\xi$ found in \cite{Bazavov:2016nty} turns out to be significantly smaller than its tree-level value in (\ref{gamma}).

\begin{figure}[!tb]
\centering
\includegraphics[width = 0.55\textwidth]{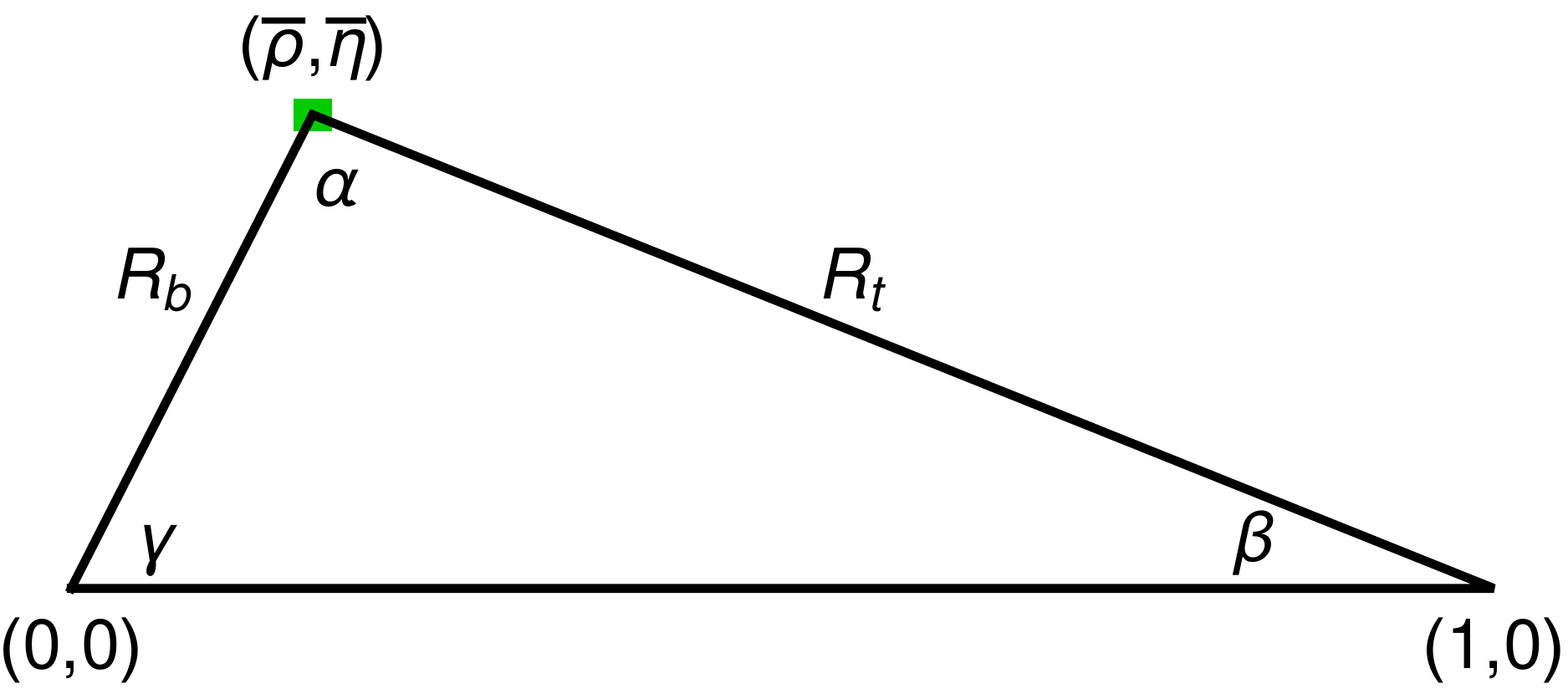}
 \caption{\it Universal Unitarity Triangle 2016. The green square at the apex 
of the UUT shows that the uncertainties in this triangle are impressively small.}\label{UUTa}
\end{figure}

Now,
\be\label{simple}
\vtd=\vus\vcb R_t\,,\qquad \vts=\eta_R\vcb~
\ee
with $R_t$ being one of the sides of the unitarity triangle (see Fig.~\ref{UUTa})
and  
\be
\eta_R=1 -\vus\xi\sqrt{\frac{\Delta M_d}{\Delta M_s}}\sqrt{\frac{m_{B_s}}{m_{B_d}}}\cos\beta+\frac{\lambda^2}{2}+\ord(\lambda^4) = {0.9825}\,,
\ee
where we have used
\be\label{beta}
\beta=(21.85\pm0.67)^\circ \,
\ee
obtained from 
\be\label{SpKs}
S_{\psi K_S}=\sin 2\beta =0.691 \pm 0.017 \,.
\ee

Thus using (\ref{DMSDMD}) and (\ref{simple}) 
we determine very precisely 
\be\label{RT} 
R_t= 0.741 \, \xi = {0.895\pm0.013}\,.
\ee

Having determined $\beta$ and $R_t$ we can construct the UUT shown in Fig.~\ref{UUTa}, from 
which we find 
\be
{\bar\rho=0.170\pm0.013 \,,\qquad\bar\eta= 0.333\pm0.011}\,.
\ee
{We observe that the UUT in Fig.~\ref{UUTa} differs significantly from the UT obtained in global fits \cite{Charles:2015gya,Bona:2006ah}, with the latter exhibiting smaller $\bar\rho$ and larger $\bar\eta$ values.}

\begin{figure}[!tb]
\centering
\includegraphics[width = 0.55\textwidth]{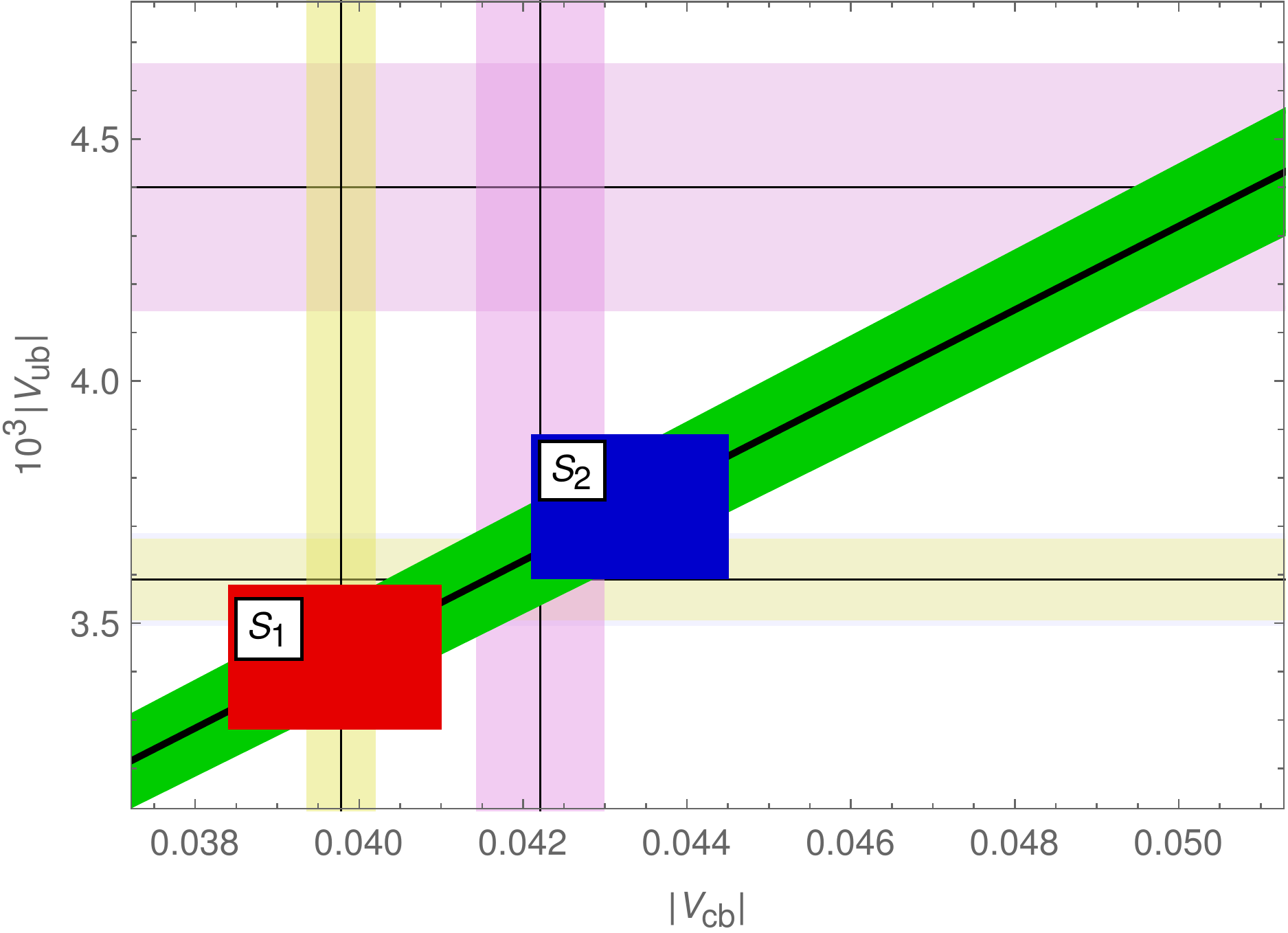}
 \caption{\it $\vub$ versus $\vcb$ in CMFV (green) compared with the tree-level 
exclusive (yellow) and inclusive (violet) determinations. The squares are our results in $S_1$ (red) and $S_2$ (blue).}\label{VubVcbplot}
\end{figure}

Subsequently, using the 
relation
\be\label{VUBG}
 R_b=\left(1-\frac{\lambda^2}{2}\right)\frac{1}{\lambda}
\left| \frac{V_{ub}}{V_{cb}} \right|=
\sqrt{1+R_t^2-2 R_t\cos\beta}
\ee
allows a very precise determination of  the ratio 
\be\label{vubvcb}
\frac{\vub}{\vcb}= {0.0864\pm 0.0025} \,.
\ee
This implies, as shown in Fig.~\ref{VubVcbplot}, a strict correlation between $\vub$ and $\vcb$
 that can be compared with the tree-level determinations of both CKM elements, 
also shown in this plot. The exclusive  determinations have been summarized 
in \cite{DeTar:2015orc} and are given as follows
\be\label{excl}
\vcb_\text{excl}=(39.78\pm0.42)\cdot 10^{-3},\qquad \vub_\text{excl}=(3.59\pm0.09)\cdot 10^{-3}.
\ee
They are based on \cite{Lattice:2015rga,Lattice:2015tia,Glattauer:2015teq,Bazavov:2016nty,Aaij:2015bfa}.
The inclusive ones are summarized well in 
\cite{Alberti:2014yda,Gambino:2016fdy}.
\be\label{incl}
\vcb_\text{incl}=(42.21\pm0.78)\cdot 10^{-3},\qquad \vub_\text{incl}=(4.40\pm0.25)\cdot 10^{-3}.
\ee
{We note that after the recent Belle data on $B\to D\ell\nu_l$ \cite{Glattauer:2015teq}, the exclusive and inclusive values of $\vcb$ are closer to each other 
than in the past. On the other hand in the case of $\vub$ there is a very significant difference. But the inclusive value for $\vub$ 
implies new CP phases in order to accomodate the data on $S_{\psi K_S}$ and 
consequently the CMFV framework selects the exclusive value of $\vub$ as we 
will see below.}\footnote{We thank Paolo Gambino and Ruth van de Water for private communication on this topic.}

 We observe that within the CMFV framework only special combinations  of these two CKM elements are allowed. The {\it red} and {\it blue} squares represent the 
ranges obtained in the strategies $S_1$ and $S_2$, respectively, as explained below and summarized in 
Table~\ref{tab:CKM}. We observe significant tensions both between the results  in $S_1$ and $S_2$ and {also between them and the inclusive tree-level determination of $\vub$. On the other hand  
the exclusive determination of $\vub$ accompanied by the inclusive 
one for $\vcb$ gives  $\vub/\vcb=0.0850\pm 0.0026$, very close to
the result in (\ref{vubvcb}).} However the separate values of $\vub$ and $\vcb$ 
in (\ref{excl}) and (\ref{incl}) used to obtain this result are not compatible with our findings in $S_1$, implying problems with
$\Delta M_{s,d}$ as we will see below.

\begin{figure}[!tb]
\centering
\includegraphics[width = 0.55\textwidth]{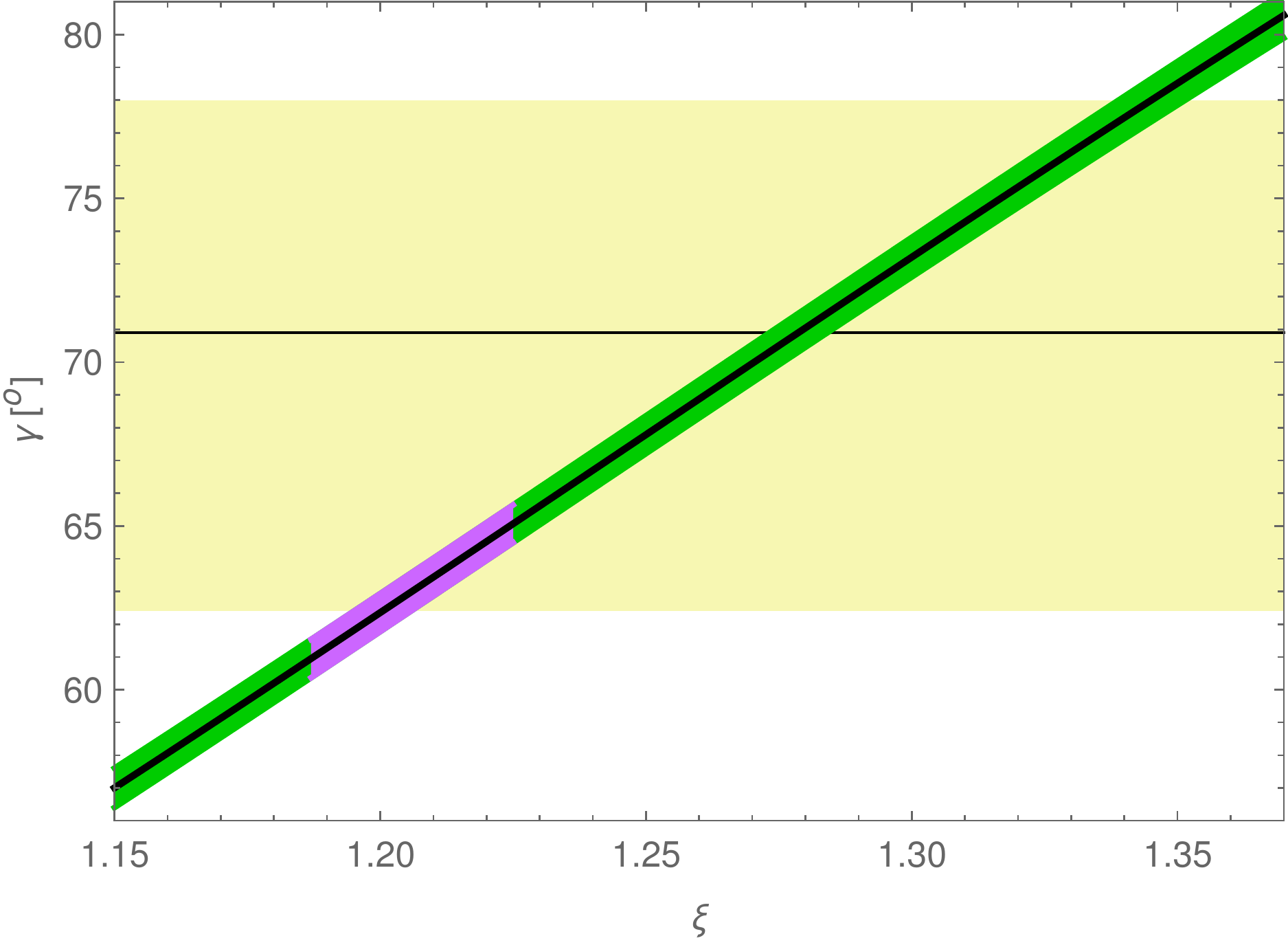}
 \caption{\it $\gamma$ versus $\xi$ for $S_{\psi K_S}=0.691\pm0.017$.
The violet range corresponds to the new lattice determination of $\xi$ in (\ref{xi}), and the yellow range displays the tree-level determination of $\gamma$ (\ref{gamma}).}\label{fig:gammavsxi}
\end{figure}

Returning to the issue of the origin of the difference between (\ref{DMSDMD})
and (\ref{treeDMSDMD}), the new lattice results \cite{Bazavov:2016nty} have  important implications on 
the angle $\gamma$  in the UUT that can be determined by means of
\be\label{VUBG2}
\cot\gamma=\frac{1-R_t\cos\beta}{R_t\sin\beta}\,.
\ee
With the very precise value of $\xi$ and consequently $R_t$ we can
precisely determine  the angle $\gamma$  
independently of the values of $S(v)$, $\vub$ and $\vcb$. In Fig.~\ref{fig:gammavsxi} we show $\gamma$ as a function of $\xi$ 
from which we extract 
\be\label{newgamma}
{\gamma=(63.0\pm 2.1)^\circ}\,,
\ee
below its central value from tree-level decays in (\ref{gamma}), 
and with an uncertainty that is by more than a factor of three smaller.
We will use this value in what follows.  
 We note that
the uncertainty due to $S_{\psi K_S}$ is very 
small. In order to appreciate this result one can read off the plot in 
Fig.~\ref{fig:gammavsxi}  that the old range of $\xi=1.27\pm 0.06$ 
corresponds to  $\gamma=(70\pm 6)^\circ$.

Finally, from (\ref{beta}) and (\ref{newgamma}) we determine the angle $\alpha$ 
in the unitarity triangle 
\be\label{alpha}
{\alpha= (95.1\pm 2.2)^\circ}\,.
\ee

It should be emphasized that the results in (\ref{beta}), (\ref{RT}), (\ref{vubvcb}), (\ref{newgamma}) and (\ref{alpha}) are independent of $S(v)$ and therefore valid for all CMFV models.

\boldmath
\subsection{$S_1$: Upper Bounds on $\vts$, $\vtd$, $\vcb$, $\vub$ and $\varepsilon_K$}
\unboldmath

Returning to (\ref{DMS}) and (\ref{DMD}), we note that the overall factors on the r.h.s.\ equal the central experimental values 
of $\Delta M_s$ and $\Delta M_d$, respectively. We can therefore read off from these 
formulae the central values of $\vts$ and $\vtd$ corresponding to the 
lattice results in (\ref{Kronfeld}). Including the uncertainties in the latter 
formula and taking into account the inequality (\ref{BBBOUND}) 
we find the {\it maximal} values of $\vts$ and $\vtd$ in the CMFV models 
that are consistent with the data on $\Delta M_s$ and $\Delta M_d$ 
\be\label{CMFVVTSVTD}
{
\vts_\text{max}=(39.0 \pm 1.3)\cdot 10^{-3}, \qquad \vtd_\text{max} = (8.00\pm 0.29)\cdot 10^{-3}.
}
\ee
It should be noted
that
\be\label{General}
{
\vts=39.0\cdot 10^{-3}\,\sqrt{\frac{2.322}{S(v)}}\,, \qquad
\vtd= 8.00\cdot 10^{-3}\,\sqrt{\frac{2.322}{S(v)}},
}
\ee
where we suppressed the errors given in (\ref{CMFVVTSVTD}). Thus 
the bounds in (\ref{CMFVVTSVTD}) are saturated in the SM. The results within the SM are in excellent agreement with those obtained in \cite{Bazavov:2016nty}. Yet, here we also
stress that these are upper bounds in CMFV models. Therefore, the tension 
between the values of these CKM elements extracted from $\Delta M_{s,d}$ 
and their tree-level determinations found in \cite{Bazavov:2016nty} within the SM is larger
in any other CMFV model. Interestingly the values of $\vtd$ and $\vtd/\vts$ 
extracted from the rare semi-leptonic decays $B\to\pi\mu^+\mu^-$ and 
 $B\to K\mu^+\mu^-$ agree with the ones in (\ref{General}) and (\ref{DMSDMD}), 
respectively \cite{Du:2015tda}: 
\be\label{DMSDMDrare}
\frac{\vtd}{\vts}=0.201(20)\,,\qquad \vts=35.7(1.5)\cdot 10^{-3}\,\,\qquad 
\vtd=7.45(69)\cdot 10^{-3}\,.
\ee
 For $\vts$, the values are found to be even smaller than in  (\ref{General}).
{However this determination of CKM parameters still suffers from large uncertainties.  We refer to \cite{Bazavov:2016nty} for a more detailed 
comparison of rare semileptonic $B$-decays with $B_{s,d}$ mixing results and 
the relevant references.}

\begin{figure}[!tb]
\centering
\includegraphics[width = 0.55\textwidth]{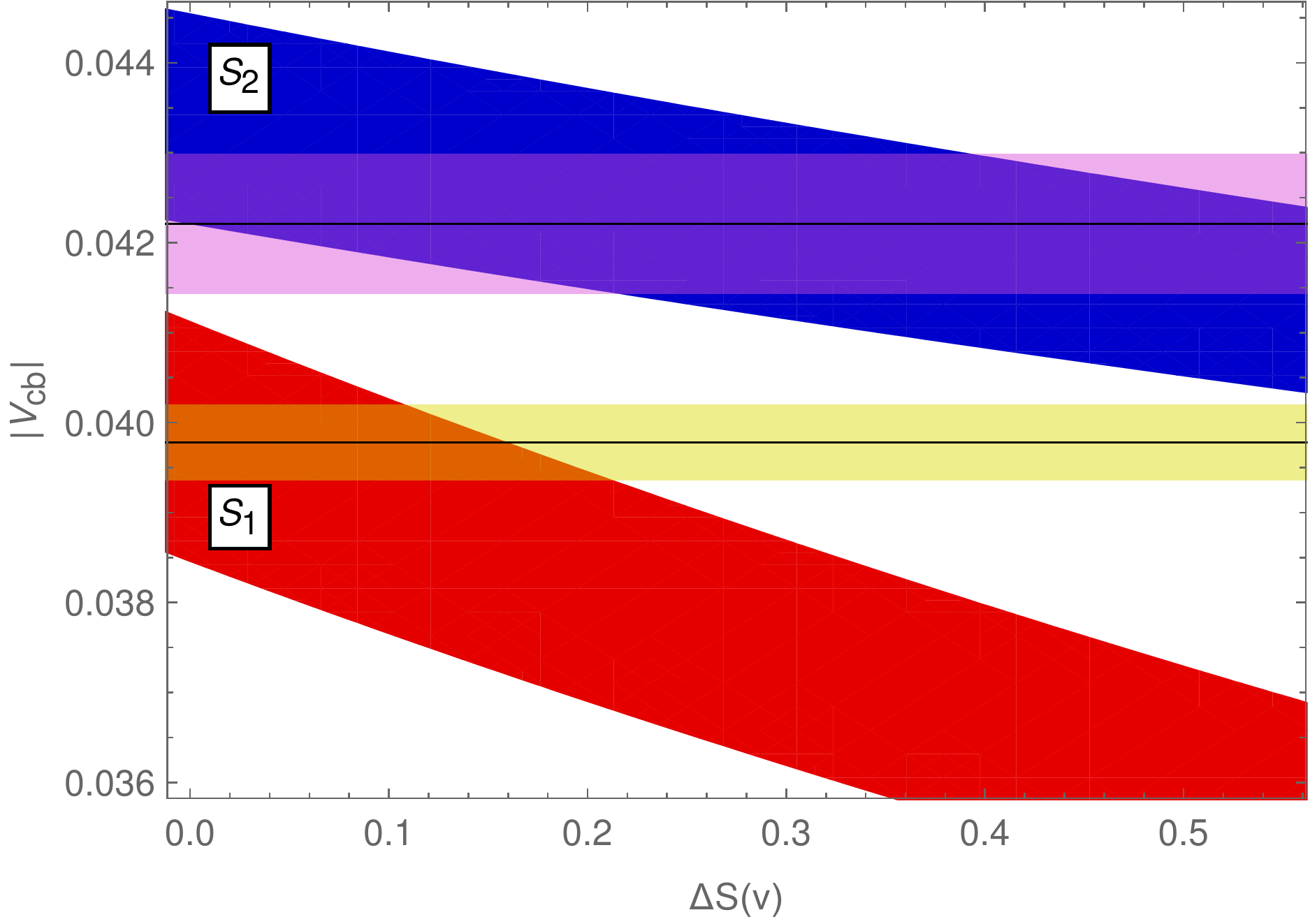}
 \caption{\it $\vcb$ versus {the flavour-universal NP contribution} $\Delta S(v)$ obtained in $S_1$ (red) and $S_2$ (blue). The horizontal bands correspond to the tree-level measurements in (\ref{excl}) (yellow)  and 
(\ref{incl}) (violet).}\label{CMFVplot}
\end{figure}

With the knowledge of $\vus$, $\vts$, $\vtd$ and $\beta$ 
we can determine $\vub$ and $\vcb$ as functions of $S(v)$ so that they 
can directly be compared with their determinations from semi-leptonic decays 
summarized in (\ref{excl}) and (\ref{incl}).
We find 
\be
{\vcb= (39.7\pm 1.3)\cdot 10^{-3}\,\sqrt{\frac{2.322}{S(v)}}, 
\qquad \vub= (3.45\pm 0.15)\cdot 10^{-3}\,\sqrt{\frac{2.322}{S(v)}}\,.}
\ee
This dependence is represented by the {\it red} band in Fig.~\ref{CMFVplot} with $\Delta S(v)$
defined by
\be
S(v)=S_0(x_t)+\Delta S(v)\,.
\ee
For illustrative purposes we also show the tree-level values in (\ref{excl}) and (\ref{incl}). 
Evidently the exclusive determinations of $\vcb$ are favoured in $S_1$. {Furthermore with increasing $\Delta S(v)$, $\vcb$ quickly drops significantly below the value in (\ref{excl}).}

Having the full CKM matrix as a function of $S(v)$, we can calculate 
 the CP-violating parameter $\varepsilon_K$. We use the usual formulae which can be found in \cite{Buras:2013raa}. It should be noted that $\varepsilon_K$ 
depends directly on 
\be
V_{ts}=-\vts\, e^{-i\beta_s},\qquad V_{td}=\vtd\, e^{-i\beta}
\ee
with $\beta_s=-1^\circ$. Consequently, the value of $\vcb$ is not needed for
this evaluation.

Now, the dominant contribution to $\varepsilon_K$ is 
proportional to 
\be\label{econstraint}
|\varepsilon_K|\propto \vts^2\vtd^2 S(v) \propto \frac{1}{S(v)},
\ee 
where we have used (\ref{General}).  Thus
with $\vts$ and $\vtd$ determined through $\Delta M_{s,d}$, the parameter 
 $\varepsilon_K$ {\it decreases} with increasing $S(v)$, in contrast to the 
analysis in which the CKM parameters are taken from tree-level decays. In that case 
$\varepsilon_K$ increases with increasing $S(v)$.

Consequently using $S_1$ we find the upper bound on $\varepsilon_K$ in CMFV models
to be 
\be\label{ebound}
{|\varepsilon_K|\le (1.64\pm0.25)\cdot 10^{-3}\,.}
\ee
We conclude that the imposition of the $\Delta M_{s,d}$ 
constraints within CMFV models implies an upper bound on 
$\varepsilon_K$, saturated in the SM, which is significantly below its experimental value given in Table~\ref{tab:input}. Therefore a non-CMFV contribution 
\be \label{DES}
{|\varepsilon_K|_\text{non-CMFV} \ge (0.59\pm 0.25)\cdot 10^{-3}}
\ee
is required, implying a discrepancy of the SM and CMFV value of $\varepsilon_K$ {with the data} 
by {$2.3\,\sigma$}. Once more we stress that this shift cannot be obtained within CMFV models without violating the constraints from $\Delta M_{s,d}$.

In Table~\ref{tab:CKM} we collect the values of the most relevant CKM parameters as well 
as the real and imaginary parts of $\lambda_t=V_{td} V^*_{ts}$. In particular 
the value of ${\rm Im}\lambda_t$ is important for the ratio $\epe$. Its 
value found in $S_1$  is lower than what has been used in the recent papers  \cite{Bai:2015nea,Buras:2015yba,Buras:2015xba,Buras:2015jaq}, thereby further decreasing
the value of $\epe$ in the SM.

\begin{table}[!tb]
\centering
{
\begin{tabular}{|c|c|c|c|c|c|c|}
\hline
 $S_i$ & $\vts$ & $\vtd$ & $\vcb$ & $\vub$ &${\rm Im}\lambda_t$ & ${\rm Re}\lambda_t$        \\
\hline
$S_1$ &$39.0(13)$   & $8.00(29)$ & $39.7(13)$& $3.43(15)$ & $ 1.21(8)$& $-2.88(19)$\\
$S_2$ & $42.6(11)$  &$8.73(26) $ & $43.3(11) $& $3.74(14)$& $1.44(7)$ & $-3.42(18)$ \\
 \hline
\end{tabular}}
\caption{\it Upper bounds on CKM elements in units of $10^{-3}$ and of $\lambda_t$ in units of $10^{-4}$ obtained using strategies 
$S_1$ and $S_2$ as explained in the text. We set $S(v)=S_0(x_t)$. 
}\label{tab:CKM}
\end{table}

\boldmath
\subsection{$S_2$: Lower Bounds on $\Delta M_{s,d}$}
\unboldmath
The strategy $S_2$ uses the construction of the UUT as outlined above, but then
instead of using $\Delta M_s$ for the complete extraction of the CKM elements, the experimental value of $\varepsilon_K$ is used as input. Taking the lower bound in (\ref{BBBOUND}) into account, this strategy again implies   {\it upper  bounds} on $\vts$, $\vtd$, $\vcb$ and $\vub$. However this time their $S(v)$ dependence differs from the one in (\ref{General}), as seen in the case of $\vcb$ 
in Fig.~\ref{CMFVplot}, where $S_2$ is represented by the {\it blue} band. {The weaker $S(v)$ dependence in $S_2$, together with the higher $\vcb$ values, is another proof that the tension between $\varepsilon_K$ and $\Delta M_{s,d}$ cannot be removed within the CMFV framework and is in fact smallest in the SM limit.}

In order to understand this weaker dependence of $\vcb$ on $S(v)$  we use the formula for  $\vcb$ extracted from $\varepsilon_K$ 
that has been derived in \cite{Buras:2013raa}. We recall
it here for convenience\footnote{We replaced $v(\eta_{cc},\eta_{ct})$ by $\tilde v(\eta_{cc},\eta_{ct})$ in order to distinguish it from the argument in $S(v)$.}
\be\label{vcb1}
\vcb=\frac{\tilde v(\eta_{cc},\eta_{ct})}{\sqrt{\xi S(v)}}\sqrt{\sqrt{1+h(\eta_{cc},\eta_{ct}) S(v)}-1}\approx 
\frac{\tilde v(\eta_{cc},\eta_{ct})}{\sqrt{\xi}}\left[\frac{h(\eta_{cc},\eta_{ct})}{S(v)}\right]^{1/4}, 
\ee
where for the central values of the QCD corrections $\eta_{cc}$ and $\eta_{ct}$ in 
Table~\ref{tab:input} one finds
\be
\quad \tilde v(\eta_{cc},\eta_{ct}) = 0.0282,\quad 
h(\eta_{cc},\eta_{ct})=24.83~.
\ee
Values of  $\tilde v(\eta_{cc},\eta_{ct})$ and 
$h(\eta_{cc},\eta_{ct})$ in the full range of  $\eta_{cc}$ and $\eta_{ct}$ can be found in Table 3 of \cite{Buras:2013raa}.

Inserting (\ref{vcb1}) into (\ref{simple}) we find  
\be\label{General2}
\vts\propto \frac{1}{S(v)^{1/4}}, \qquad \vtd\propto \frac{1}{S(v)^{1/4}}\,
\ee
and consequently from  (\ref{DMS}) and (\ref{DMD})
\be
\Delta M_s \propto \sqrt{S(v)}, \qquad \Delta M_d \propto \sqrt{S(v)}\,.
\ee
Therefore,  with (\ref{BBBOUND}), we find  {\it lower bounds} 
on $\Delta M_s$ and $\Delta M_d$ that are significantly larger than the data
\be\label{Msdbounds}
{\Delta M_s \ge (21.1\pm 1.8)\text{ps}^{-1}, \qquad \Delta M_d \ge (0.600\pm0.064) \text{ps}^{-1}\, .}
\ee
Consequently, our results for $\Delta M_s$ and $\Delta M_d$ in the SM differ from their experimental values by $1.8\sigma$ and $1.5\sigma$, respectively. This {difference} increases for other CMFV models.  On the other hand, as seen in Fig.~\ref{CMFVplot},
 the value of $\vcb$ in $S_2$ is fully compatible with its tree-level determination from inclusive decays, but for small $\Delta S(v)$  larger than its exclusive determination.

The ratio of the central values of $\Delta M_{s,d}$ obtained by us 
\be\label{RCMFV}
\left(\frac{\Delta M_s}{\Delta M_d}\right)^{\rm CMFV}= {35.2}
\ee
 perfectly agrees with the data as this ratio is used in $S_1$ and $S_2$ as experimental input in our analysis. The error on this 
ratio calculated directly from (\ref{Msdbounds}) is spurious as
we impose this ratio from experiment and the true 
error is negligible. Only when one individually calculates $\Delta M_s$ and 
$\Delta M_d$ with $\vcb$ extracted from $\varepsilon_K$, the errors in (\ref{Msdbounds}) are found. However they are correlated and cancel in the ratio.

{On the other hand, using the tree-level determination of the CKM 
matrix, the authors of \cite{Bazavov:2016nty} find in the SM} 
\be\label{FMILC}
(\Delta M_s)^{\rm SM} = (19.6\pm 1.6 )\text{ps}^{-1}, \qquad (\Delta M_d)^{\rm SM} = (0.630\pm 0.069 ) \text{ps}^{-1}\,
\ee
and 
\be
\left(\frac{\Delta M_s}{\Delta M_d}\right)^{\rm SM}= 31.2\pm 1.7 \,.
\ee
{Compared with (\ref{RCMFV}), this shows the inconsistency between the tree-level determination of the CKM matrix and $\Delta F=2$ processes in CMFV models.}

 In Table~\ref{tab:CKM} we  compare the results for the CKM elements obtained in $S_2$ with the ones found using $S_1$. In both cases we use the SM value for $S(v)$, as it allows to obtain 
values of $\varepsilon_K$ in $S_1$ and of $\Delta M_{s,d}$ in $S_2$ closest 
to the data. But as we can see, the  values of the CKM elements obtained
 in $S_2$ differ by much from the corresponding ones in $S_1$, and in particular favour the inclusive determination of $\vcb$. Also the value of ${\rm Im}\lambda_t$ is larger, 
{however it differs only by a few percent from} the one used in recent {calculations} of 
$\epe$ \cite{Bai:2015nea,Buras:2015yba,Buras:2015xba,Buras:2015jaq}.

\begin{figure}[!tb]
\centering
\includegraphics[width = 0.49\textwidth]{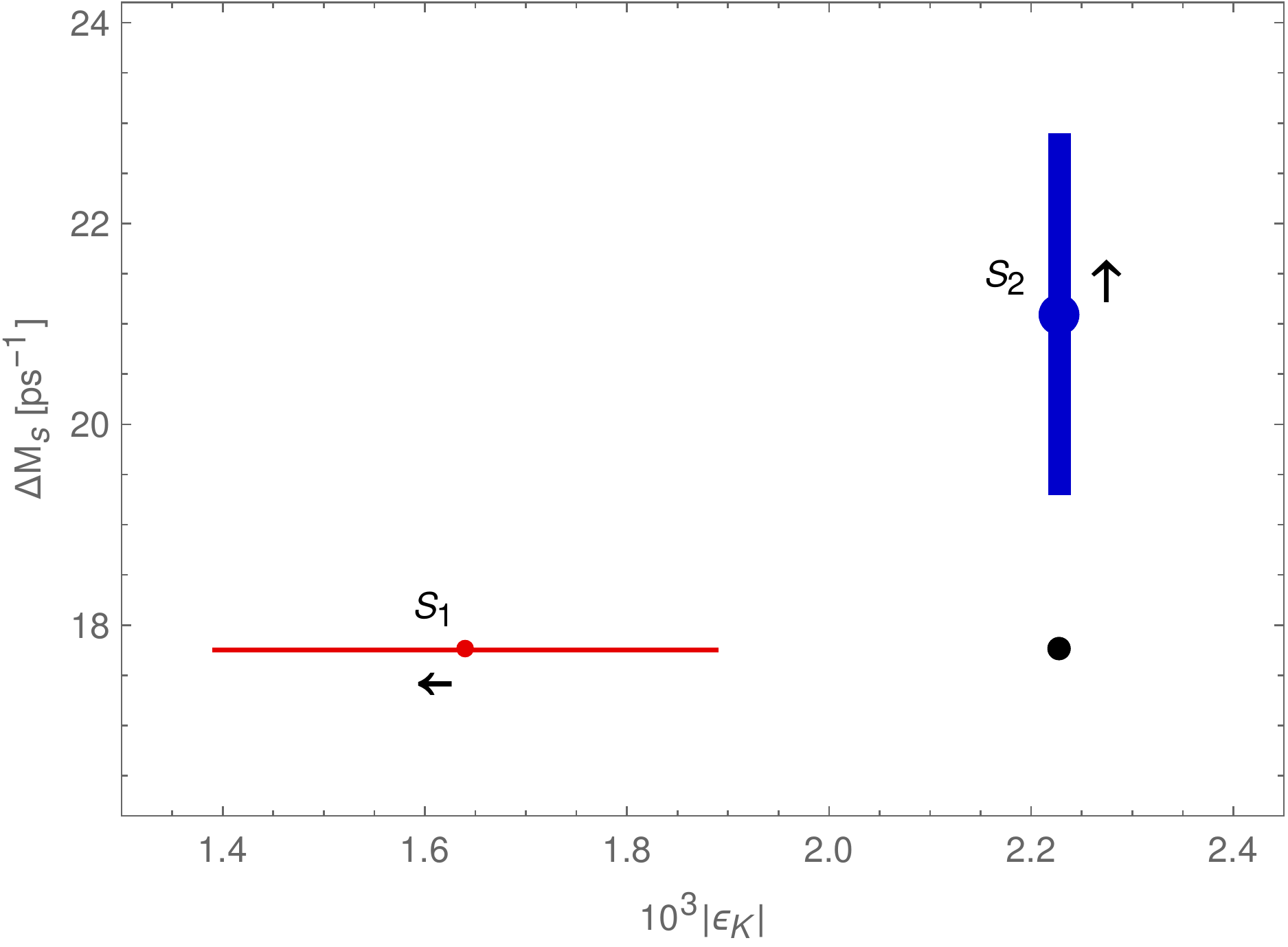}\hfill
\includegraphics[width = 0.49\textwidth]{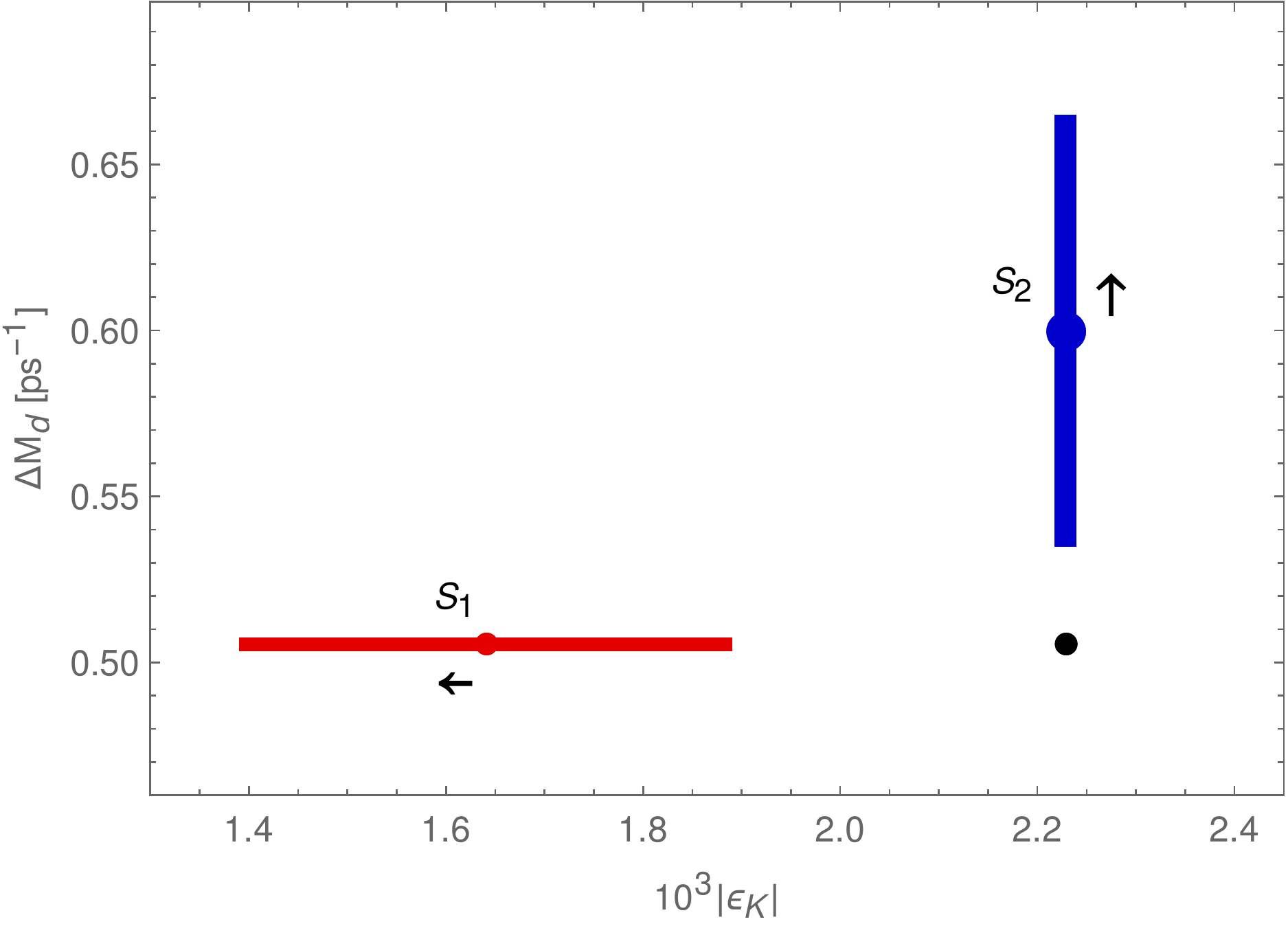}
 \caption{\it  $\Delta M_{s,d}$ and  $\varepsilon_K$ obtained from the strategies
$S_1$ and $S_2$ for $S(v)=S_0(x_t)$, at which the upper bound on $\varepsilon_K$ in $S_1$ and lower bound on  $\Delta M_{s,d}$ in $S_2$ are obtained.
The arrows show how the red and blue regions move with increasing $S(v)$. {The black dot represents the experimental values.}}\label{eKvsDMsdplots}
\end{figure}

We conclude therefore, as already indicated by the analysis in  \cite{Buras:2013raa}, that it is impossible within CMFV 
models to {obtain a simultaneous agreement of $\Delta M_{s,d}$ and $\varepsilon_K$ with the data.} 
The improved lattice results in 
(\ref{Kronfeld}) and (\ref{xi}) allow to exhibit this difficulty 
stronger. In the context of the strategies $S_1$ and $S_2$, the tension 
between $\Delta M_{d,s}$ and $\varepsilon_K$ is summarized by the plots
of $\Delta M_{s,d}$ vs. $\varepsilon_K$ in Fig.~\ref{eKvsDMsdplots}. Note that these 
plots differ from  the known plots of $\Delta M_{s,d}$ vs. $\varepsilon_K$ 
in CMFV models (see e.g. Fig.~5 in  \cite{Buras:2013ooa}). {In the latter plot} the CKM parameters were taken from tree-level decays, and varying $S(v)$ increased both 
 $\Delta M_{s,d}$ and  $\varepsilon_K$ in a correlated manner. Even if 
the physics in those plots and in the plots in Fig.~\ref{eKvsDMsdplots} is the same, presently the accuracy of the outcome of strategies $S_1$ and $S_2$ shown in 
Fig.~\ref{eKvsDMsdplots} is higher.

The problems with CMFV models encountered here could be anticipated on the 
basis of the first three rows of Table 2 from \cite{Buras:2013raa}, which we recall  in Table~\ref{tab:CMFVpred}. In that paper a different strategy has been used and various quantities have been predicted in CMFV models as functions of $S(v)$ and $\gamma$. As the first three 
columns correspond to $\gamma=63^\circ$ and $\xi=1.204$, very close to 
the values of these quantities found in the present paper, there is a clear message from Table~\ref{tab:CMFVpred}. The predicted values 
of  $F_{B_s}\sqrt{\hat B_{B_s}}$ and $F_{B_d} \sqrt{\hat B_{B_d}}$ are 
significantly below their recent values from \cite{Bazavov:2016nty} in (\ref{Kronfeld}). Moreover, 
with increasing $S(v)$ there is a clear disagreement between {the values of these parameters favoured by CMFV}  
and the values in (\ref{Kronfeld}). 
We {also refer} to the plots in  Fig.~4 of  \cite{Buras:2013raa}, where the 
correlations between $\vcb$ and $F_{B_d} \sqrt{\hat B_{B_d}}$  and 
between $\vcb$ and $F_{B_s} \sqrt{\hat B_{B_s}}$  implied by CMFV have been 
shown. Already in 2013 there was some tension between the grey regions {in that figure}
representing the 2013 lattice values and the CMFV predictions. With the 2016 lattice values in 
(\ref{Kronfeld}), the grey areas {shrunk} and moved away from the {values favoured by CMFV}. Other problems of CMFV seen from the point of view of the 
strategy in  \cite{Buras:2013raa} are listed in Section 3 of that paper.

\begin{table}[!tb]
\centering
\begin{tabular}{|c|c||c|c|c|c|c|c|c|c|}
\hline
 $S(v)$  & $\gamma$ & $\vcb$ & $\vub$ & $\vtd$&  $\vts$ & $F_{B_s}\sqrt{\hat B_{B_s}}$ & 
$F_{B_d} \sqrt{\hat B_{B_d}}$ & $\xi$ &  $\mathcal{B}(B^+\to \tau^+\nu)$\\
\hline
\hline
  \parbox[0pt][1.6em][c]{0cm}{} $2.31$ & $63^\circ$ & $43.6$ & $3.69$  & $8.79$ & $42.8$ & $252.7$ &$210.0$ & $1.204$ &  $0.822$\\
 \parbox[0pt][1.6em][c]{0cm}{}$2.5$ & $63^\circ$ & $42.8$& $3.63$ & $8.64$ & $42.1$ & $247.1$ & $205.3$ &$1.204$ &  $0.794$\\
 \parbox[0pt][1.6em][c]{0cm}{}$2.7$ &$63^\circ$ & $42.1$ & $3.56$ & $8.49$ &  $41.4$  & 
$241.8$ & $200.9$ & $1.204$ &  $0.768$\\
 \hline
\end{tabular}
\caption{\it CMFV predictions for various quantities as functions of 
$S(v)$ and $\gamma$. The four elements of the CKM matrix are in units of $10^{-3}$,  
 $F_{B_s} \sqrt{\hat B_{B_s}}$ and $F_{B_d} \sqrt{\hat B_{B_d}}$ in MeV and  $\mathcal{B}(B^+\to \tau^+\nu)$ in units of $10^{-4}$. From \cite{Buras:2013raa}.
}\label{tab:CMFVpred}
\end{table}

\boldmath
\section{Implications for Rare $K$ and $B$ Decays in the SM}\label{sec:3}
\unboldmath
In the previous section we have determined the full CKM matrix using {in turn} the strategies $S_1$ and $S_2$. It is
interesting to {determine} the impact of these new determinations on the branching
ratios of the rare decays $\kpn$, $\klpn$ and $B_{s,d}\to\mu^+\mu^-$ { within the SM}. To this 
end we use for $\kpn$ and $\klpn$ the parametric formulae derived in 
\cite{Buras:2015qea} which we recall here for completeness
\bea
    \mathcal{B}(\kpn)_\text{SM} &=& (8.39 \pm 0.30) \cdot 10^{-11} \cdot
    \bigg[\frac{\left|V_{cb}\right|}{40.7\cdot 10^{-3}}\bigg]^{2.8}
    \bigg[\frac{\gamma}{73.2^\circ}\bigg]^{0.74},\label{kplusApprox}
\\
    \mathcal{B}(\klpn)_\text{SM} &=& (3.36 \pm 0.05) \cdot 10^{-11} \cdot
    \bigg[\frac{\left|V_{ub}\right|}{3.88\cdot 10^{-3}}\bigg]^2
    \bigg[\frac{\left|V_{cb}\right|}{40.7\cdot 10^{-3}}\bigg]^2
    \bigg[\frac{\sin(\gamma)}{\sin(73.2^\circ)}\bigg]^{2}.\qquad
\eea

For $B_{s}\to\mu^+\mu^-$ we use the formula from
 \cite{Bobeth:2013uxa}, slightly modified in  \cite{Buras:2013ooa}
\be
 \overline{\mathcal{B}}(B_{s}\to\mu^+\mu^-)_{\rm SM} = (3.65\pm0.06)\cdot 10^{-9} \left[\frac{m_t(m_t)}{163.5 \gev}\right]^{3.02}\left[\frac{\alpha_s(M_Z)}{0.1184}\right]^{0.032} R_s
\label{BRtheoRpar}
\ee
where
\be
\label{Rs}
R_s=
\left[\frac{F_{B_s}}{227.7\mev}\right]^2
\left[\frac{\tau_{B_s}}{1.516 {\rm ps}}\right]\left[\frac{0.938}{r(y_s)}\right]
\left[\frac{|V_{ts}|}{41.5\cdot 10^{-3}}\right]^2.
\ee
The ``bar'' in (\ref{BRtheoRpar}) indicates that $\Delta\Gamma_s$ 
effects \cite{DescotesGenon:2011pb,DeBruyn:2012wj,DeBruyn:2012wk} have been taken into account through
 \be\label{rys}
r(y_s)=1-y_s, \qquad 	y_s\equiv\tau_{B_s}\frac{\Delta\Gamma_s}{2}
=0.062\pm0.005.
\ee

For $B_d\to \mu^+\mu^-$ one finds  \cite{Bobeth:2013uxa}
\begin{equation}
\mathcal{B}(B_{d}\to\mu^+\mu^-)_{\rm SM} = (1.06\pm 0.02)\cdot 10^{-10} 
\left[\frac{m_t(m_t)}{163.5 \gev}\right]^{3.02}\left[\frac{\alpha_s(M_Z)}{0.1184}\right]^{0.032} R_d
\label{BRtheoRpard}
\end{equation}
where
\be
R_d=\left[\frac{F_{B_d}}{190.5\mev}\right]^2
\left[\frac{\tau_{B_d}}{1.519 {\rm ps}}\right]\left[\frac{|V_{td}|}{8.8\cdot 10^{-3}}\right]^2.
\ee

In Table~\ref{tab:rare} we collect the results for the four branching ratios in the SM 
obtained using the strategies $S_1$ and $S_2$ for the determination of the CKM parameters and other updated parameters collected in Table~\ref{tab:input}. We observe significant differences in these two determinations, which 
 gives another support for the tension between $\Delta M_{s,d}$ and 
$\varepsilon_K$ in {the SM, holding more generally in CMFV models}.

\begin{table}[!tb]
\centering
{
\begin{tabular}{|c|c|c|c|c|}
\hline
$S_i$  & $ {\mathcal{B}}(\kpn)$ & $ {\mathcal{B}}(\klpn)$ & $\overline{\mathcal{B}}(B_{s}\to\mu^+\mu^-)$ & $\mathcal{B}(B_{d}\to\mu^+\mu^-)$\\
\hline
$S_1$ &$7.00(71)\cdot 10^{-11}$   & $2.16(25)\cdot 10^{-11}$  &$3.23(24)\cdot 10^{-9}$ &$0.90(8)\cdot 10^{-10}$ \\
$S_2$ &$8.93(74)\cdot 10^{-11}$    &$3.06(30)\cdot 10^{-11}$   &$3.85(24)\cdot 10^{-9}$ & $1.08(8)\cdot 10^{-10}$\\
 \hline
\end{tabular}
}
\caption{\it SM predictions for rare decay branching ratios using the strategies 
$S_1$ and $S_2$, as explained in the text.
}\label{tab:rare}
\end{table}

Our results for $B_{s,d}\to \mu^+\mu^-$ should be compared with the
{results of the combined analysis of CMS and LHCb data} \cite{CMS:2014xfa}
\be\label{LHCb2}
\overline{\mathcal{B}}(B_{s}\to\mu^+\mu^-) = (2.8^{+0.7}_{-0.6}) \cdot 10^{-9}, 
\quad
\mathcal{B}(B_{d}\to\mu^+\mu^-) =(3.9^{+1.6}_{-1.4})\cdot 10^{-10}. 
\ee
We observe that in $S_1$ the SM prediction for $B_s\to\mu^+\mu^-$ 
is rather close to the data, while in the case of $S_2$ it is visibly larger.
 {The most recent result from ATLAS \cite{Aaboud:2016ire}, 
\be
\overline{\mathcal{B}}(B_s\to\mu^+\mu^-)=(0.9^{+1.1}_{-0.9})\times 10^{-9}\,,
\ee  
while not accurate, also favours a low value for $\overline{\mathcal{B}}(B_s\to\mu^+\mu^-)$.}

Finally, in view of the improved lattice determinations of the parameters 
$\hat B_{B_s}$ and $\hat B_{B_d}$ \cite{Bazavov:2016nty}
\be
\hat B_{B_s}=1.44\pm 0.10,\qquad \hat B_{B_d}=1.38\pm 0.13
\ee
it is tempting to calculate the $B_{s,d}\to\mu^+\mu⁻$ branching ratios 
by {normalizing them to $\Delta M_{s,d}$} \cite{Buras:2003td}. This eliminates not only the dependence 
on the CKM parameters and weak decay constants, but also reduces the dependence 
on $m_t$. Neglecting the tiny uncertainties in $\eta_B$, $\alpha_s$ and $\tau_{B_q}$
we find the very accurate expressions
\bea
 \overline{\mathcal{B}}(B_{s}\to\mu^+\mu^-)_{\rm SM} &=& (3.24\pm0.05)\cdot 10^{-9} \left[\frac{1.44}{\hat B_{B_s}}\right]\left[\frac{0.938}{r(y_s)}\right]\left[\frac{m_t(m_t)}{163.5 \gev}\right]^{1.5}
\label{BRsAJB}
\,,\\
 {\mathcal{B}}(B_{d}\to\mu^+\mu^-)_{\rm SM} &=& (0.91\pm 0.02)\cdot 10^{-10} \left[\frac{1.38}{\hat B_{B_d}}\right] \left[\frac{m_t(m_t)}{163.5 \gev}\right]^{1.5}\,.
\label{BRdAJB}
\eea

These expressions apply only to the SM and $S_1$, {where the experimental values of $\Delta M_{s,d}$ are used to determine the CKM matrix}. We then find 
\be\label{AJBBsd}
\overline{\mathcal{B}}(B_{s}\to\mu^+\mu^-)_\text{SM} = (3.24\pm 0.20) \cdot 10^{-9}, 
\quad
\mathcal{B}(B_{d}\to\mu^+\mu^-)_\text{SM} =(0.91\pm 0.08)\cdot 10^{-10}. 
\ee
These results agree perfectly with the ones in Table~\ref{tab:rare}. 
This is not surprising  because in $S_1$ the constraint from  $\Delta M_{s,d}$ has been imposed and the authors of \cite{Bazavov:2016nty} extracted the values of 
$\hat B_{B_q}$ from their results in (\ref{Kronfeld}) and $F_{B_q}$ in Table~\ref{tab:input}.
{The outcome of this exercise will be more illuminating once independent and more precise lattice determinations 
of the $\hat B_{B_{s,d}}$ parameters become available.
In addition}, the derived formulae (\ref{BRsAJB}) and (\ref{BRdAJB})
are much simpler than the ones in (\ref{BRtheoRpar}) and (\ref{BRtheoRpard}), 
respectively. They allow in no time to calculate the branching ratios in 
question in terms of  $\hat B_{B_s}$, $\hat B_{B_d}$,  $\Delta\Gamma_s$ and 
$m_t$.

\section{Beyond CMFV}\label{sec:4}

Our analysis of CMFV models signals the violation  of flavour universality in 
the function $S(v)$, {signalling the presence of new sources of flavour and CP-violation and/or
new 
operators contributing to $\Delta F=2$ transitions beyond the SM $(V-A)\otimes (V-A)$ ones.\footnote{In a more general formulation of MFV new operators could 
be present \cite{D'Ambrosio:2002ex}.} For simplicity we will here restrict ourselves to  solutions in which only SM operators 
are present.}

{A fully general and very convenient solution} in this case is just to consider instead of the
 flavour universal function $S(v)$ three functions
\be
S_i=|S_i| e^{i\varphi_i}, \qquad  i=K,s,d \,.
\ee

It is evident that with two free parameters in each meson system it is possible 
to obtain an agreement with the data on $\Delta F=2$ observables. The simplest models of this type are models with tree-level $Z^\prime$ and $Z$ exchanges analysed in detail in \cite{Buras:2012jb}. The flavour violating couplings in these 
models are complex numbers (two free parameters) and can be chosen in such a manner that any problems of CMFV models in $\Delta F=2$ processes are removed by properly choosing these couplings. Effectively the observables in (\ref{great5}) are simply used to  find these parameters or equivalently $S_i$. The test 
of these scenarios is only offered through the correlations with $\Delta F=1$ processes, that is rare $K$ or $B_{s,d}$ 
decays, which in these simple models  involve the same couplings.
The analysis in  \cite{Buras:2012jb} then shows that when constraints from 
 $\Delta F=1$ processes are taken into account it is easier to obtain an 
agreement with the data for $\Delta F=2$ processes in the case of $Z^\prime$ models than models with tree-level  $Z$ exchanges.

Here we would like to discuss only the models with a minimally broken $U(2)^3$ flavour symmetry \cite{Barbieri:2011ci,Barbieri:2012uh} 
which are more constrained. In these models, as discussed in detail in \cite{Buras:2012sd}, in addition to the unitary CKM matrix 
one has 
\be\label{G1}
S_K=r_K S_0(x_t),\qquad r_K\ge 1
\ee
and 
\be
 |S_d|=|S_s|=r_B S_0(x_t), \quad \varphi_{d}=\varphi_{s}\equiv\varphi_{\rm new}
\ee
with $r_B$ being a real parameter which could be larger or smaller than unity.
The important difference from the CMFV scenario is that it cannot be tested 
without invoking tree-level determinations of at least some elements of the 
CKM matrix.  The main features of this scenario are:
\begin{itemize}
\item
No correlation between the $K$ and $B_{s,d}$ systems, so that the tension between 
$\varepsilon_K$ and $\Delta M_{s,d}$ is absent in these models. 
\item
However as $r_K\ge 1$, finding one day $\varepsilon_K$ in the SM to be larger 
than the data would exclude
this scenario. {Presently such a situation seems rather unlikely.}
\item
$S_d \equiv S_s$ are complex functions and $r_B$ can be larger or smaller than 
unity. Consequently,  through interference with the SM contributions, 
$\Delta M_{s,d}$ can 
be suppressed or enhanced {as needed}.
\item
With the new phase $\varphi_{\rm new}$ and $r_B$ not bounded from below there is
more freedom than in the CMFV scenario.
\end{itemize}

However, due to the equality $S_d=S_s$ there are two important implications that 
can be tested.

The first one is the  CMFV relation \cite{Buras:2012sd}
\be\label{CMFV3}
\left(\frac{\Delta M_d}{\Delta M_s}\right)_{{\rm M}U(2)^3}=
\left(\frac{\Delta M_d}{\Delta M_s}\right)_{\rm CMFV}=
\left(\frac{\Delta M_d}{\Delta M_s}\right)_{\rm SM}=
\frac{m_{B_d}}{m_{B_s}}
\frac{1}{\xi^2}
\left|\frac{V_{td}}{V_{ts}}\right|^2.
\end{equation}
from which one can obtain the ratio $\vtd/\vts$ as done already in 
section \ref{sec:2}, see (\ref{DMSDMD}), which can be {compared 
with its tree-level determination}. As stated before, the tree-level determination of this ratio, quoted in (\ref{treeDMSDMD}), is significantly larger, and 
consequently M$U(2)^3$ models have the same difficulty here as CMFV models. 
Yet, a firm conclusion will only be  reached after the result in 
 (\ref{treeDMSDMD}) will be {superseded} by a more precise tree-level determination
of the angle $\gamma$.

The second one is the correlation between the two CP asymmetries that results 
from the equality of NP phases in
\begin{equation}
S_{\psi K_S} = \sin(2\beta+2\varphi_{\rm new})\,, \qquad
S_{\psi\phi} =  \sin(2|\beta_s|-2\varphi_{\rm new})\,,\qquad 
({\rm M}U(2)^3)\,.
\label{U21}
\end{equation}
{As $\beta_s$ is very small in the SM, a precise measurement of $S_{\psi\phi}$ 
determines $\varphi_{\rm new}$. From the measured value of 
$S_{\psi K_S}$ we then obtain $\beta$. The latter value can be compared with the one obtained from the tree-level determination of $\vub/\vcb$ and either $R_t$ or the tree-level determination of $\gamma$.}
However $\beta$ is strongly correlated  with $\vub/\vcb$, with very weak dependence 
on  $\gamma$ and $R_t$. Therefore eventually (\ref{U21}) 
implies a triple correlation between\cite{Buras:2012sd}
\be\label{Triple}
S_{\psi K_S}\,,\qquad S_{\psi\phi}\,,\qquad \frac{\vub}{\vcb}\,,
\ee
which provides another important test 
of the M$U(2)^3$ 
scenario once the three observables will be known precisely. 

In summary, M$U(2)^3$ models face the new lattice data  better than 
CMFV, but similar to the latter models have difficulties with the value of 
$\gamma$ and of the ratio $\vtd/\vts$ being significantly below their 
tree-level determinations. 

Concerning more complicated models like the Littlest Higgs model with T-parity \cite{Blanke:2006sb,Blanke:2015wba} or 
331 models \cite{Buras:2015kwd},  it is clear that the new lattice data 
has an impact on the allowed ranges of new parameters. However such a study is beyond the scope of our paper\footnote{{For a recent analysis in the 331 models
see \cite{Buras:2016dxz}.}}.

\section{Conclusions}\label{sec:5}

In this paper we have determined the {universal unitarity triangle (UUT) of constrained minimal flavour violation (CMFV)} models. {We then derived the full CKM matrix, using either the experimental value of $\Delta M_s$ or of $|\eps_K|$ as input. The recently improved values of the hadronic matrix elements in (\ref{Kronfeld})
and (\ref{xi}) \cite{Bazavov:2016nty} have been crucial for this study.} In contrast to many  analyses in the literature, we avoided tree-level determinations of $\vub$ and $\vcb$.

The main messages from this analysis are as follows:
\begin{itemize}
\item
The extracted angle $\gamma$ in the UUT is already  known {precisely} and is somewhat smaller than its 
tree-level determination. This is a direct consequence of the small value of 
$\xi$ in (\ref{xi}). In turn the ratio $\vtd/\vts$ also turns out to be 
smaller than its tree-level determination, as already pointed out in \cite{Bazavov:2016nty}.
\item
The precise relation between $\vub$ and $\vcb$ obtained 
by us in (\ref{vubvcb}) provides another test of CMFV. See Fig.~\ref{VubVcbplot}.
\item
Requiring CMFV to reproduce the data for $\Delta M_{s,d}$ (strategy $S_1$), we find that low values of $\vub$ and $\vcb$ are favoured, in agreement with their exclusive determinations. More importantly we derived an upper bound on $|\varepsilon_K|$ that is 
significantly below the data.
\item 
Requiring CMFV to reproduce the data for $\varepsilon_K$ (strategy $S_2$), we find a  higher  value of $\vub$, still consistent with exclusive determinations, but  $\vcb$ significantly higher than in $S_1$ and in agreement with its inclusive  determination. The derived lower bounds on $\Delta M_{s,d}$   are then significantly above the data.
\item
The tension between $\varepsilon_K$ and $\Delta M_{s,d}$ in CMFV models 
with {either $|\varepsilon_K|$ being too small or 
$\Delta M_{s,d}$ being too large}
cannot be removed by varying $S(v)$. This would {only} be possible, as stressed in 
\cite{Buras:2013raa}, if the values in (\ref{Kronfeld}) turned out to be significantly 
smaller and $\xi$ larger than in (\ref{xi}). With the present 
values of these parameters, the SM performs best among all CMFV models, 
even if, as seen in Fig.~\ref{eKvsDMsdplots}, it falls short in properly describing  the $\Delta F=2$ data.
\item
The {inconsistency of $\Delta M_{d,s}$ and $\varepsilon_K$ in the SM and
CMFV is} also signalled by rather different predictions for
rare decay branching  ratios obtained using strategies $S_1$ and $S_2$. See 
Section~\ref{sec:3} and Table~\ref{tab:rare}.
\item
As the correlation between $\varepsilon_K$ 
and $\Delta M_{s,d}$ is broken in models with $U(2)^3$ flavour symmetry, 
 these models perform better than CMFV models. {Still} the correlation between
$\Delta M_s$ and $\Delta M_d$, that is of CMFV type, predicted by these
models is in conflict with the tree-level determinations already pointed out in \cite{Bazavov:2016nty}
within the SM. See (\ref{DMSDMD}) and (\ref{treeDMSDMD}). 
\end{itemize}

Our analysis of CMFV models shows that they fail to properly describe  the existing data on $\Delta F=2$ 
observables simultaneously
and implies thereby {the presence of} either new sources of flavour violation and/or new operators. Several models 
analysed in the literature like $Z^\prime$ models, 331 models, or the Littlest Higgs model with T-parity (LHT)
could help in bringing the theory to agree with the data. {In fact in the 
case of 331 models it was demonstrated recently in \cite{Buras:2016dxz} that 
the tensions pointed by us here could be removed in these models. We expect that
this is also the case in the LHT model, firm conclusions would however require a dedicated study.}

Certainly, further improvements on the hadronic matrix elements from lattice QCD 
and {on the} tree-level determinations of $\vub$, $\vcb$ and $\gamma$ will 
sharpen the prediction for the size of required NP contributions to $\Delta F=2$ observables, 
thereby selecting models which could bring the theory to agree with experimental data.  In particular finding {the value of $\gamma$ from tree-level 
determinations in the ballpark of $70^\circ$} would imply the violation of 
the CMFV relation (\ref{CMFV3}). On the other hand resolving the discrepancy 
between exclusive and inclusive tree-level determinations of $\vub$ in 
favour of the latter, would indicate the presence of new CP-violating phases 
affecting $S_{\psi K_S}$. Moreover, the correlations of $\Delta F=2$ transitions with rare $K$ and $B_{s,d}$ decays and $\epe$ could 
eventually give us a deeper insight into the NP at short distance scales that is responsible for the anomalies indicated by the new lattice data, as reviewed 
in  \cite{Buras:2013ooa} and recently stressed in \cite{Buras:2015jaq}.

{\bf Acknowledgements}\\
We thank Aida El-Khadra and  Andreas Kronfeld  for many illuminating discussions and information on the progress in their lattice calculations  and in particular for sharing their new results with us prior to publication.
The research of AJB was fully financed and done in the context of the ERC Advanced Grant project ``FLAVOUR'' (267104) and was partially supported by the DFG
cluster of excellence ``Origin and Structure of the Universe''.

\bibliographystyle{JHEP}
\bibliography{allrefs}
\end{document}